# Artificial and self-assembled pinning centers in Ba(Fe$_{1-x}$Co$_x$)$_2$As$_2$ thin films as a route to very high current density.

C. Tarantini[1], S. Lee[2], F. Kametani[1], J. Jiang[1], J.D. Weiss[1], J. Jaroszynski[1], C.M. Folkman[2], E.E. Hellstrom[1], C. B. Eom[2], D.C. Larbalestier[1]

[1] Applied Superconductivity Center, National High Magnetic Field Laboratory, Florida State University, Tallahassee FL 32310, USA

[2] Department of Materials Science and Engineering, University of Wisconsin-Madison, Madison, WI 53706, USA

We report on the superior vortex pinning of single and multilayer Ba(Fe$_{1-x}$Co$_x$)$_2$As$_2$ thin films with self-assembled *c*-axis and artificially introduced *ab*-plane pins. Ba(Fe$_{1-x}$Co$_x$)$_2$As$_2$ can accept a very high density of pins (15-20 vol%) without $T_c$ suppression. The matching field is greater than 12 T, producing a significant enhancement of the critical current density $J_c$, an almost isotropic $J_c\,(\theta,20\text{T}) > 10^5$ A/cm$^2$, and global pinning force density $F_p$ of ~ 50 GN/m$^3$. This scenario strongly differs from the high temperature cuprates where the addition of pins without $T_c$ suppression is limited to 2-4 vol%, leading to small $H_{Irr}$ enhancements and improved $J_c$ only below 3-5 Tesla.

## I. INTRODUCTION

The recently discovered Fe-based superconductors (FBS) exhibit intrinsic properties, like high critical temperature $T_c$,[1] large upper critical field $H_{c2}$ and relatively low anisotropy that generate great interest,[2] which is amplified by their complexity, their intriguing superconducting mechanisms, and their similarity to cuprate high temperature superconductors.[3] FBS have also shown high intragrain critical current density $J_c$,[4] irreversibility field $H_{Irr}$ close to $H_{c2}$,[2] and great possibility of improving their $J_c$ by introducing effective vortex pinning centers. These discoveries deserve attention. Co-doped $BaFe_2As_2$ (Ba122) thin films have been grown by several groups [(Ref.5,6)] and we demonstrated that vertically-aligned, self-assembled $BaFeO_2$ (BFO) nanorods (NR) could be introduced without suppressing $T_c$.[7,8] These nanorods act as strong $c$-axis correlated pins which enhance $J_c(H//c)$ above $J_c(H//ab)$, inverting the intrinsic material anisotropy,[7] because the nanorod diameter is comparable to $2\xi$ ($\xi$ being the superconducting coherence length). $c$-axis pinning has been subsequently found in different films, enabling, so far, pinning force density $F_p$ of about 30 $GN/m^3$ at 4 K and 12 T in the best field configuration ($H//c$).[9] Ba122 appears to be unique among the high temperature superconductors because it can accept a much higher density of pinning centers than in $YBa_2Cu_3O_{7-x}$ (YBCO).[10,11,12]

In this paper we will present a comprehensive study of the pinning properties of ameliorated single- and multi-layered Co-doped Ba122 thin films investigated in a wide temperature range and in high magnetic field up to 45T. We select four Ba122 films grown by pulsed laser deposition (PLD) with different microstructures in order to explore the pinning tunability afforded by combined artificial and self-assembled pins following an approach similar to what previously studied in YBCO case.[10-12] We will show the high effectiveness of these strongly correlated pins in enhancing the in-field performances with respect to the previously reported results.[7,9] In particular we report an increase of $F_p$, exceeding 50 GN/m$^3$, a decrease of the $J_c$ anisotropy with an almost flat angular dependence at 20 T and a $J_c$(4.2K,20T) value well above 0.1MA/cm$^2$.

## II. EXPERIMENTAL

The four Ba122 films studied in this paper were grown by PLD as described in Refs. 5 and 6 using two Ba(Fe$_{0.92}$Co$_{0.08}$)$_2$As$_2$ targets. The two targets were synthesized from different starting reactants: in the first case pure elements were employed, whereas the second target was prepared using pre-reacted Ba$_3$As$_2$ as the barium source. Both targets were heat treated at 1120°C for 12 hours and processed in the hot isostatic press. Because of a larger amount of unreacted Ba present in the first case, oxidation is more likely to occur producing a high oxygen content (HOC) target. In contrast, the second synthesis technique results in a more phase-pure material that minimizes the oxidation producing a low oxygen content (LOC) target. As a consequence, these two targets generate a different density of BFO nanorods in the films.[8] A third undoped Ba-122 target (un-HOC) was prepared by the same synthesis route as the HOC target. The so-obtained target was used for the preparation of multilayer films together with the LOC Co-doped target. Single layer films grown with the HOC target generated a high density of self-assembled BFO pins as deposited on (001)-oriented (La,Sr)(Al,Ta)O$_3$ substrates with an intermediate template of 50 or 100 unit cells (u.c.) of SrTiO$_3$ (named HOC-S50 and HOC-S100, respectively).[5] Two films were grown with the LOC target on 100 u.c. SrTiO$_3$/(La,Sr)(Al,Ta)O$_3$: one was a single layer film (LOC-S), which acted as a reference sample with low pin density while the second contained artificially introduced multilayers (LOC-M) with alternating layers of 13.3 nm of Co-doped Ba122 from the LOC target and ~3.3 nm of undoped Ba122 from the un-HOC target (total thickness ~400 nm), as reported elsewhere.[13] The film structures are summarized in Table I. The same growth conditions were used for the HOC, LOC, and un-HOC targets. The zero-resistance $T_c$ of the samples ranges between 21.0 and 22.8 K with resistive transition widths < 0.9 K.

We performed high field transport measurements at the National High Magnetic Field Laboratory in the 35 and 45 T magnets and in a Quantum Design 16 T-PPMS in order to study the pinning properties over a wide range of H-T phase diagram. $J_c$ was determined by four-contact measurements using a 1µV/cm criterion. The nature and density of the introduced defects were studied by using a JEOL JEM2011 transmission electron microscope (TEM).

## III. RESULTS

Figure 1 reveals the complex microstructure of HOC-S100. The cross-sectional TEM image of Fig. 1(a) shows the *c*-axis BFO nanorods (NR), previously reported,[7,8] as well as additional nanoparticle (NP) arrays arranged along the *ab*-planes with *c*-axis spacing of about 18 nm. The higher magnification image (Fig. 1(b)) shows that both nanoparticles and nanorods have a 4-5 nm diameter. The nanorods, which in earlier studies were continuous from the buffer layer to the top surface,[7,8] here show some discontinuity but still maintain long lengths. From the planar view of HOC-S100 (Fig. 1(c)) we estimated an average spacing of ~12.5 nm for the randomly distributed thin nanorods, corresponding to a matching field $B_\phi$ of ~13.2 T. The formation of nanoparticles, which were not present in the previous work,[7] is likely related to a different amount of oxygen that generates the BFO secondary phase. The HOC target used in this study has more oxygen than in Ref. 7, as confirmed by the larger nanorods density, and the formation of additional nanoparticles is likely related to the necessity to incorporate more oxygen without inducing excessive stress in the superconducting matrix. A similar structure was also found in HOC-S50 but with larger NP array spacing (~26 nm) and NR separation (~14.5 nm, $B_\phi$ ~ 9.8 T). In both samples we also observed that some thin nanorods merge, forming 20 nm-diameter columnar defects. A rough estimate of the combined NR and NP volume fraction is 12 and 16 vol% for HOC-S50 and HOC-S100, respectively, to which the large columnar defects contribute an additional 4.5-5.5 vol%. Because of their low density, these larger defects may not be very effective as pins (density ~100 times smaller than the thin nanorods) but they do reduce the superconducting cross-section, potentially reducing the $J_c$ deduced from transport $I_c$ measurements.

Measurements performed at 12-16 K up to 16T (Figure 2) show that HOC-S50 has a larger self- and low field $J_c$ than HOC-S100. However HOC-S100 performs better in higher field, showing $H_{Irr}^{//ab}$(16K)~15T compared to 13T in HOC-S50 ($H_{Irr}$ defined as $J_c(H_{Irr})=10^2 A/cm^2$). The opposing low- and high-field behaviors are likely related to the high density of defects that have the dual effect of blocking current at low $H$ but adding effective pins at high field. For $H//c$, both HOC films show much stronger pinning compared to $H//ab$ with $J_c(H//c)$ exceeding $J_c(H//ab)$ up to 10-11 T at 12 K.

In order to study the effectiveness of the different defects in the low-temperature (i.e. weak thermal fluctuation), very high-field regime, $J_c$ and $F_p=J_c\times\mu_0 H$ at 4.2 K up to 35 T are reported in Fig. 3(a)-(b) for the HOC samples. The data on LOC-S, the film with the lowest defect concentration, are also shown for comparison. For $H//c$, the sample microstructure clearly manifests itself in the pinning properties and $H_{Irr}$ increases from ~34 T for low-pinning LOC-S to over 40 T for high-pinning HOC-S100. The $F_p$ maximum increases from ~39 GN/m$^3$ in LOC-S to 47-53 GN/m$^3$ in the HOC samples (20-36% increase). The most striking evidence of the relation between increased $F_{p,Max}$ and the presence of nanorods is that the position of the $F_p(H//c)$ maxima (Fig.3a) almost corresponds to $B_\phi$ of the two HOC films. Moreover the drop of $J_c$ with increasing field markedly increases above $B_\phi$ (it is even more visible on a linear scale, not shown). For $H//ab$ (Fig. 3(b)) LOC-S and HOC-S100 have a similar $J_c$ and $F_p$ magnitude above 5T with $F_{p,Max} > 42$ GN/m$^3$, whereas in HOC-S50 they are significantly lower. Because both nanorods and nanoparticles can similarly contribute as pinning centers for $H//ab$, $B_\phi$ is not easy to determine but a conservative estimate can be attempted, which we do as follows. In Figure 4 the low and high magnification TEM images are shown together with a possible vortex site diagram. The layers with the nanoparticle arrays actually have two possible sources of pinning, the nanoparticles themselves and $c$-axis aligned nanorods that cross these layers. Because of this large defect concentration (nanoparticles plus nanorods) we assume that the vortices preferentially sit in these layers and, within these layers, both nanoparticles and nanorods pin the vortices (strongly pinned vortices marked with red symbols in Figure 4(c)). Other possible vortex sites could be taken into account (marked with green symbols in Figure 4(c)) but, considering that in these locations only the nanorods are present, we assume their pinning effectiveness is minor and we neglect them. The estimated $B_\phi$ values for H//$ab$ are ~6 T for HOC-S50 and ~12 T for HOC-S100. These values explain the slight bump in $F_p$ in those field regions (Fig. 3b). Figure 3(c) shows that $J_c(\theta)$ of HOC-S100 is clearly enhanced compared to LOC-S with an increase along the $c$-axis of 50, 70 and 90% at 15, 20 and 25 T. What is remarkable here is that HOC-S100 retains a weak $J_c$ angular dependence up to very high field and that $J_c(\theta)$ is still almost constant and well above $1.5 \times 10^5$ A/cm$^2$ at 20 T. Clearly the combined effects of nanorods and nanoparticles produce a significant improvement of $J_c$ at every angle, especially along the $c$-axis because of the high density and long length of the nanorods. Moreover, the similar in-field $J_c(H//ab)$ of HOC-S100 and LOC-S, despite their difference in $J_c$(s.f.), suggests that the $ab$-arranged nanoparticles give some additional contribution to $J_c$ for $H//ab$, in addition to providing isotropic random pinning.[14]

A potentially more controlled approach to artificially introduce pinning centers is to alternate superconducting and non-superconducting layers by multilayer deposition. In our case, the natural candidate for the non-superconducting material is the undoped parent compound Ba122, which is a poor metal. The TEM images in Fig. 5 point out that the multilayer deposition did not produce continuous undoped layers, but rather a distribution of flat or round *ab*-plane nanoprecipitates, which is actually advantageous since these defects can act as effective pins without compromising the continuity of the superconducting matrix. The average layer separation along the *c*-axis is about 16-17 nm, consistent with the deposition conditions. Figure 5 reveals three possible sources of pinning: flat precipitates (3-4 nm thick and 8-30 nm wide), 2-4 nm round-shaped nanoparticles and short nanorods. A portion of the nanoparticles is generated by the deposition of the undoped material and they are aligned in *ab*-layers while others are randomly distributed between these layers. Figure 5 also shows that in LOC-M there are only a few nanorods and they are short compared to the abundant and long nanorods in HOC-100 (Fig. 1 and 4). In particular, in LOC-M they appear either to be chopped by or to nucleate on the nanoparticle/flat-precipitate *ab*-layers, whose separation limits the NR length. The low nanorod density in LOC-M is due to the low oxygen content of the LOC target, since oxygen is necessary to form $BaFeO_2$.[7,8]

Figure 6 compares LOC-S and LOC-M at high temperature (12-16 K). The multilayer deposition has no negative effect on $J_c$(s.f.), and LOC-M $J_c$(H//c) shows a low-field dependence similar to LOC-S, but with a slight improvement at high field. However a clear enhancement of $J_c$(H//ab) occurs in LOC-M at low field (Fig. 6(b)), as emphasized also in the angular dependence (Fig. 7). At 12 K the enhanced region extends to ~7-8 T (Fig. 6b) and the $F_p$ curve (inset of Fig. 6) shows two separate peaks for LOC-M indicative of at least two different pinning mechanisms.

Decreasing temperature to 4.2 K, the effectiveness of the additional pinning centers changes (Fig. 8). In LOC-M, $J_c$(H//c) improves over the whole field range compared to LOC-S with the irreversibility field increasing from 34.5 T (LOC-S) to 40.5 T (LOC-M) and $F_p$(H//c) is enhanced by ~20%. For H//ab, LOC-M has better performance to over 30 T with $F_p$(H//ab) increasing from 44 to 53 $GN/m^3$. In this case a rough estimate of $B_\phi$ is 8-9 T, which explains the wide shape of the $F_p$ peak. The angular dependence of $J_c$ (Fig. 8(c)) reveals that the additional pins in LOC-M have a quite uniform effect, inducing an almost isotropic increase in $J_c(\theta)$. The superimposed pinning effects of the flat *ab*-precipitates, the short *c*-axis nanorods, and the round nanoparticles produce an almost perfectly isotropic vortex pinning landscape. It is also interesting to note how the effectiveness of different pinning mechanisms changes with temperature. At 12 K $J_c(H//c)$ is similar for LOC-S and LOC-M and the difference along *ab* is evident only at low field (Figs. 6 and 7). At 4.2 K the uniform increase of $J_c(\theta)$ for LOC-M up to high field suggests that additional pins are activated at low temperature, strengthening both the *c*-axis and *ab*-plane pinning and positively affecting the intermediate angles. Considering the microstructure of LOC-M, this extra source of pinning likely arises from the round-shaped nanoparticles. Because of their small diameter (2-4 nm), most of them are smaller than $2\xi$ at 12 K, so they are too small to oppose the thermal fluctuations. At lower temperature, where $\xi$ decreases and thermal fluctuations are suppressed, their additional contribution as random pins becomes very effective.

The data reported here clearly show a strong pinning enhancement by both *self-assembled* BFO and *artificially layered* defects. Specifically, in the HOC samples, which contain *self-assembled* defects, the diameter of the BFO nanorods and nanoparticles is comparable to $2\xi$, making the nanorods effective pins along the *c*-axis and the nanoparticles effective over a wide angular range. Moreover the *ab*-alignment of the nanoparticles provides an additional *ab*-plane pinning. This high defect density in the HOC films significantly improves the in-field performance in a wide angular range without compromising the superconducting properties of the matrix, as confirmed by their high $T_c$. However the 20 vol% of secondary phases in HOC-S100 does reduce the current-carrying cross-section and thus $J_c$(s.f.). Despite this, the combined pinning effects produce an almost isotropic $J_c$(4.2K,20T) exceeding $1.5\times10^5$ A/cm$^2$. Moreover $F_p$(4.2K) of HOC-S100 reaches 42-47 GN/m$^3$ at 20-15 T (H//*ab* and H//*c*, respectively), significantly larger than previously reported (30 GN/m$^3$).[9] An even larger $F_p$(4.2K) ~ 53GN/m$^3$ has been obtained but in the most anisotropic and so less useful HOC-S50.

In the case of *artificial defects* introduced by the multilayer deposition of doped/undoped Ba122 (LOC-M), the flat *ab*-precipitates, the round nanoparticles and the short *c*-axis nanorods generate a complex precipitate landscape that develops an almost isotropic pinning and, as a consequence, similar angular dependence $J_c(\theta)$ to that of the single layer sample (LOC-S). Interestingly, the small nanoparticles in the LOC-M seem to play a central role in enhancing the vortex pinning when the temperature decreases. In contrast to HOC-S100, the low density and short length of the nanorods and the presence of the *ab*-aligned precipitated in LOC-M make $J_c(H//ab)$ larger than $J_c(H//c)$ at high field and the $F_p$ maxima are $F_{p,\max}(H//ab)$ = 52.6 GN/m$^3$ at ~17.5 T and $F_{p,\max}(H//c)$ = 47 GN/m$^3$ at ~10 T. As a consequence, even though $J_c$(4.2K,20T) still exceeds 10$^5$ A/cm$^2$, LOC-M develops a more anisotropic $J_c$ than HOC-S100.

## IV. DISCUSSION AND CONCLUSIONS

In conclusion we showed that a surprisingly high content of non-superconductive phase, up to 20%vol., can be introduced in Ba122 thin films, either by self-assembled defect formation or by multilayer deposition. In the most isotropic sample (HOC-S100) the matching field $B_\phi$ exceeds 12 T in both the *c*-axis and *ab*-plane directions, strongly ameliorating the in-field properties and developing a highly desirable weak $J_c$ field-dependence. The irreversibility field at 4.2 K exceeds 40 T, remarkably high for a material whose $T_c$ is only 21-23 K. Moreover, the shape of the pinning force curve $F_p(H//ab)$ is characteristic of the very strong pinning seen in Nb47wt.Ti,[15] close to $h(1-h)$, where $h = H/H_{c2}$ and $h_{\max}\approx 0.5$, rather than at $h_{\max}$~0.2 as seen in Nb$_3$Sn.[16] Despite similar $T_c$ and $F_{p,\max}$ values, $F_{p,\max}$ in Nb$_3$Sn is at only 5T because sparse grain boundaries provide the effective pins while in Ba122 the high pin density pushes $F_{p,\max}$ to 10-20T. Moreover the pinning in Ba122 is exceptionally tunable compared to YBCO, which also can accept many types of secondary phase defects like $BaZrO_3$ nanorods [(Ref.10,12)] or $RE_2O_3$ nanoparticles,[11] but their volume fraction cannot exceed 2-4%vol. ($B_\phi$~3-5T) without decreasing $T_c$. Ba122 also shows an interesting temperature dependence of the pinning properties (particularly in the multilayer case) that are somehow reminiscent of what observed in YBCO where the different temperature effectiveness of weak and strong pinning was studied.[17] In our Ba122 films however, because of the high-field effectiveness of the pins, a similar investigation regarding the different pinning mechanisms cannot be carried out.

There are several possible reasons for the differences between Ba122 and YBCO related to both intrinsic material properties and chemical/structural match of the superconducting phase with the non-superconducting defects, as discussed in detail in Ref. 3. Despite the many similarities between FBS and cuprates, like layered structures, charge transfer between layers, proximity to the antiferromagnetic phase, low carrier densities, unconventional symmetries ($s^\pm$ wave and d-wave) in the pair mechanisms, and short coherence length (1-2nm) that determines the Cooper pair size, there are also important differences in the two classes of materials. The most obvious is the anisotropy $\gamma$ ($\gamma = (m_c/m_{ab})^{1/2}$, with $m_c$ and $m_{ab}$ being the effective masses in the main crystallographic directions) that is less than 2 in Ba122 but about 5-6 in YBCO. Moreover $T_c$ in FBS is only weakly affected by the introduction of defects as shown by irradiation experiments,[18,19] whereas in YBCO the $T_c$-suppression is more important.[20,18] In regard to the chemical/structural match of the superconducting phase with the non-superconducting defects, in Co-doped Ba122, both BFO and undoped Ba122 defects seem to induce little strain in the surrounding matrix, whereas in YBCO significant buckling of the *ab*-planes around the defects is observed which affects the local doping.[21] In case of large nanodots in YBCO, intergrowth can also occur producing severe bending of (00l) planes as well.[22] These important microstructural deformations combined with the larger YBCO anisotropy can induce a strong suppression of the superconducting properties around the defects making the introduction of high pin density in YBCO impossible. All of these observations support continued exploration of Co-doped Ba122 for its intrinsic materials interest and potentially too for future applications.

**Acknowledgements**

This work was supported at FSU by NSF DMR-1006584, the State of Florida and by the National High Magnetic Field Laboratory which is supported by the National Science Foundation under DMR 0084173, and the State of Florida. Work at the University of Wisconsin was supported by funding from the DOE Office of Basic Energy Sciences under award number DE-FG02-06ER46327.

**Figure captions**

FIG. 1. TEM images of HOC–S100 single layer Ba(Fe$_{0.92}$Co$_{0.08}$)$_2$As$_2$ thin film. (a)-(b), cross-section images showing the *c*-axis BFO-nanorods (NR) and the *ab*-arranged nanoparticles (NP). (c) Planar view reveals a high density of nanorods, corresponding to a matching field $B_\phi$=13.2T.

FIG. 2. $J_c$ as a function of applied field at high temperature for HOC-S50 and HOC-S100. Two field orientations are shown: (a) *H*//*c* and (b) *H*//*ab*.

FIG. 3. (a)-(b) $J_c(H)$ at 4.2 K and up to 35T with *H*//c and *H*//*ab* for three single layer Ba(Fe$_{0.92}$Co$_{0.08}$)$_2$As$_2$ thin films deposited from HOC and LOC targets. The inset shows the pinning force densities (the arrows indicate $B_\phi$). (c) $J_c(\theta)$ for LOC-S and HOC-S100 thin films (the arrows emphasize the $J_c$ enhancement).

FIG. 4 (a)-(b) Low and high magnification TEM images of HOC-S100 from Fig.1a-b showing the nanorods and the presence of nanoparticles arranged along the ab-planes and with a mean distance $\bar{d}$ in the c-direction of ~18 nm. (c) The same image of figure (b) superimposed with the possible vortex sites. Here the H//b case is depicted. The red crossed circles represent strongly pinned vortices, which intersect both nanorods and nanoparticles (considered to estimate the matching field in case of H $\perp$ c), while the green crossed circles represent weakly pinned vortices, which lie between the nanoparticle arrays and intersect just the nanorods (not considered in the matching field estimate).

FIG. 5. Cross-sectional TEM image of LOC-M, a multilayer $Ba(Fe_{0.92}Co_{0.08})_2As_2$ thin film, showing the multilayer structure and the presence of 3-4nm thick plate-like precipitates parallel to the *ab*-planes, short vertically aligned nanorods and round nanoparticles (some of them are indicated by horizontal, vertical and 45° arrows, respectively).

FIG. 6. $J_c(H)$ at high temperature for LOC-S and LOC-M for (a) *H//c* and (b) *H//ab*. The inset shows the pinning force density for *H//ab*.

FIG. 7. $J_c(\theta)$ at 12K for LOC-S and LOC-M. At low field the additional pinning along the *ab*-planes is very effective, but the differences decrease upon increasing field.

FIG. 8. (a)-(b) $J_c(H)$ at 4.2 K and up to 45T with *H//c* and *H//ab* for LOC-S and LOC-M. The inset shows the pinning force densities (the arrows indicate $B_\phi$). (c) $J_c(\theta)$ for the same samples (the arrows emphasize the $J_c$ enhancement).

**TABLE I.** Thin film structures: sample name, substrate and buffet layer information, Ba-122 targets used in the deposition (LOC, low oxygen content; HOC, high oxygen content; un-HOC, undoped- high oxygen content), structure description and thickness.

| Sample name | Substrate/Buffer Layer | Target(s) | Description and thickness |
|---|---|---|---|
| HOC-S50 | LSAT/50u.c.STO | HOC | Single-layer (~ 420nm) |
| HOC-S100 | LSAT/100u.c.STO | HOC | Single-layer (~ 420nm) |
| LOC-S | LSAT/100u.c.STO | LOC | Single-layer (~ 400nm) |
| LOC-M | LSAT/100u.c.STO | LOC + un-HOC | Multi-layer (~ 400nm) [(13.3nm Co-Ba122 + 3.3nm Ba122)×24] |

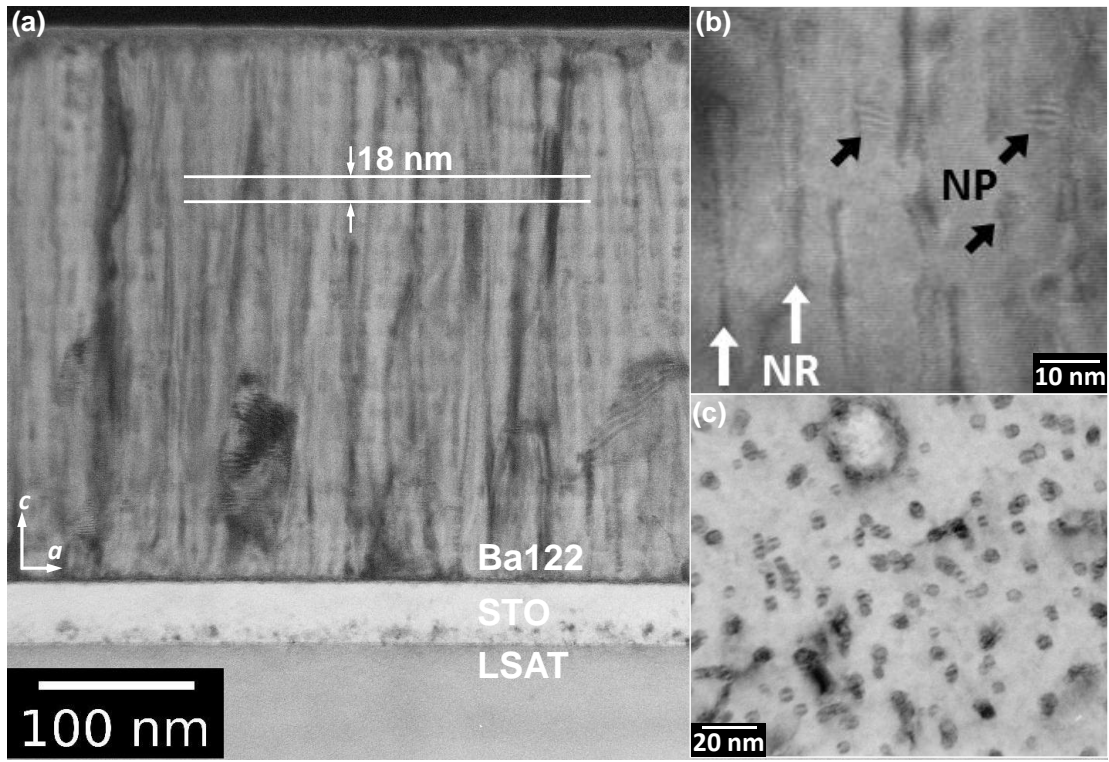

**Figure 1**

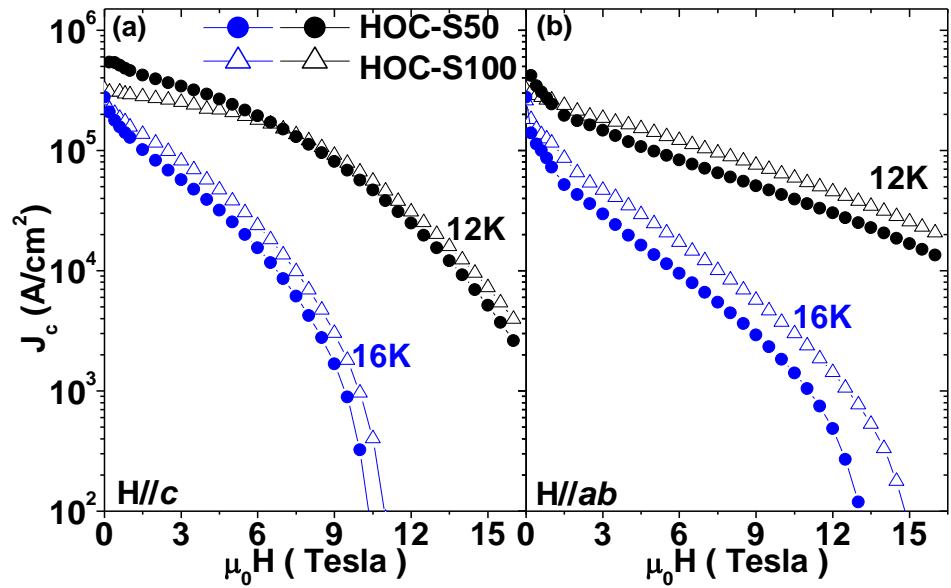

**Figure 2**

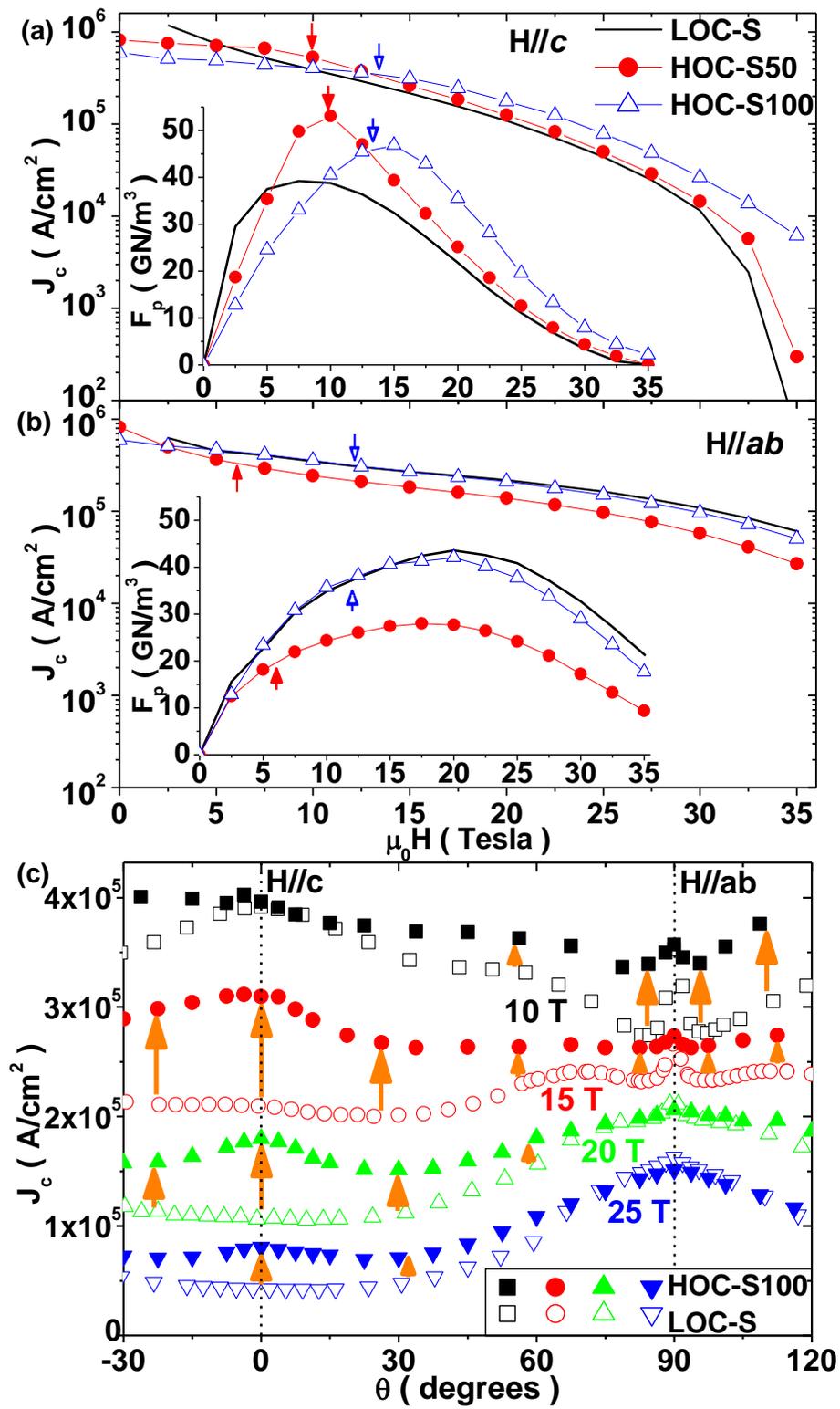

Figure 3

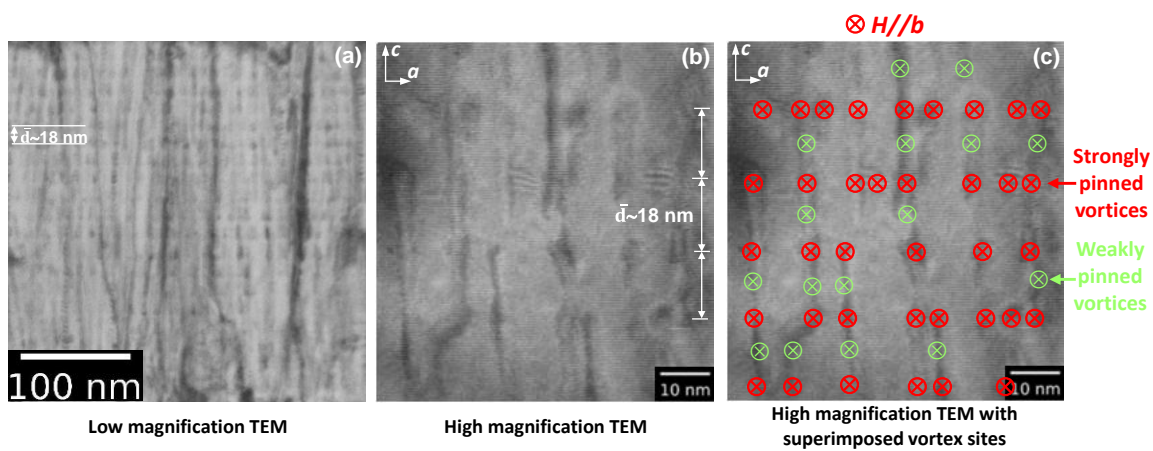

| Low magnification TEM | High magnification TEM | High magnification TEM with superimposed vortex sites |

**Figure 4**

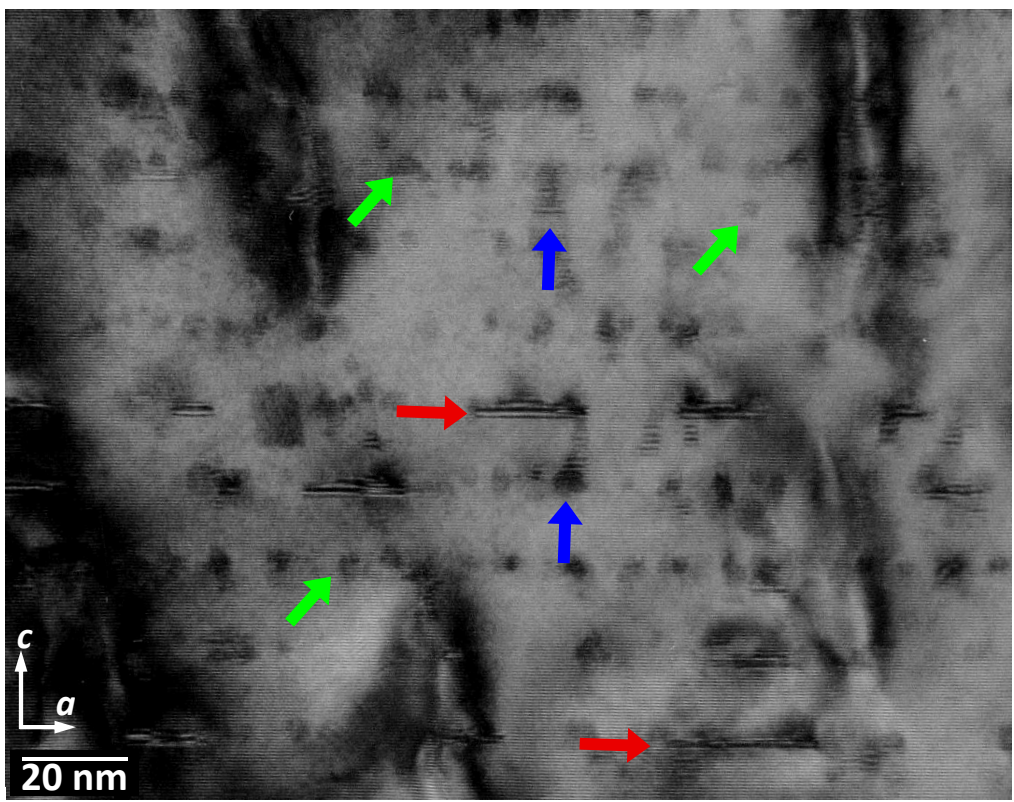

**Figure 5**

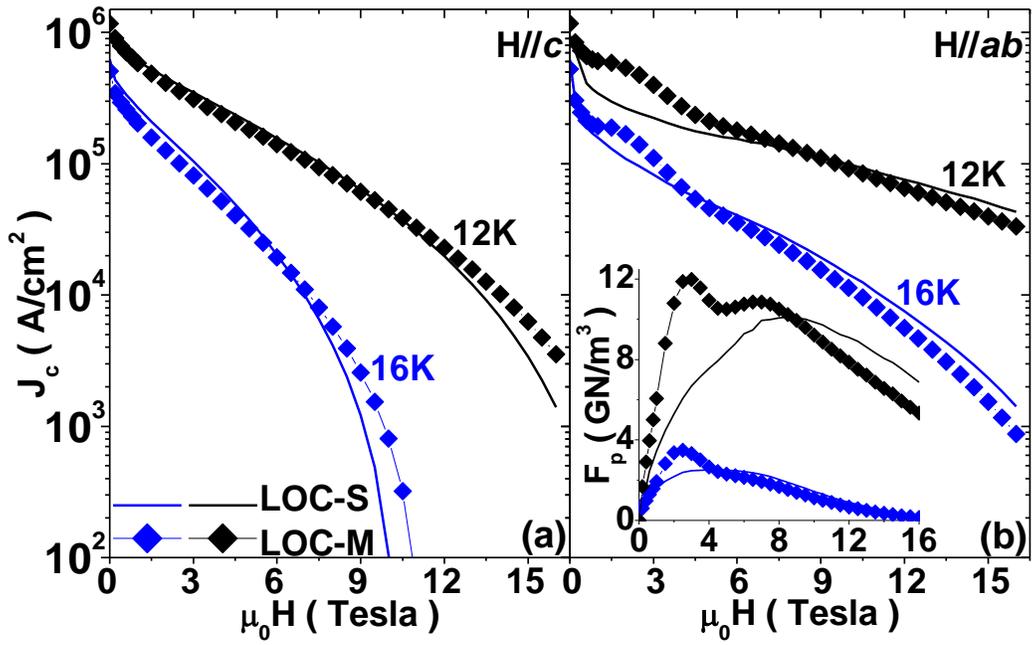

Figure 6

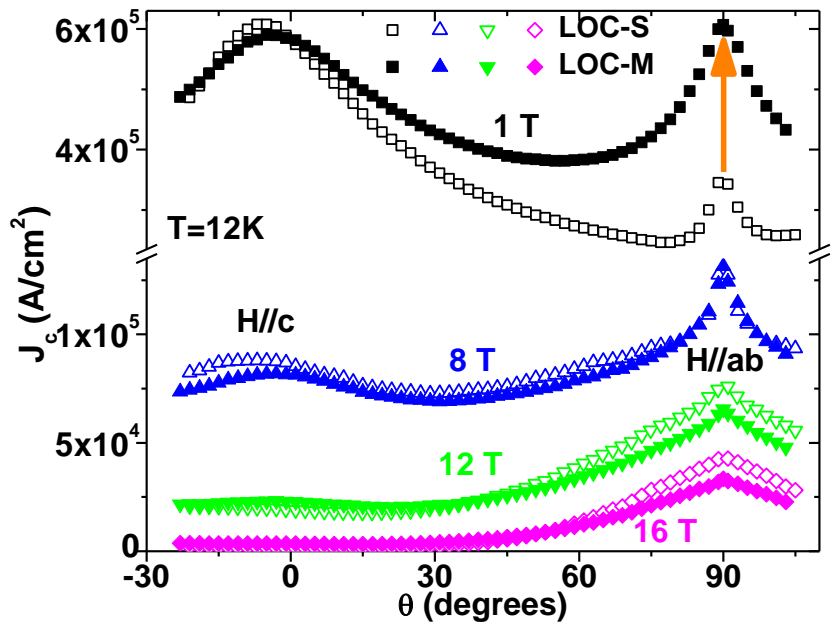

Figure 7

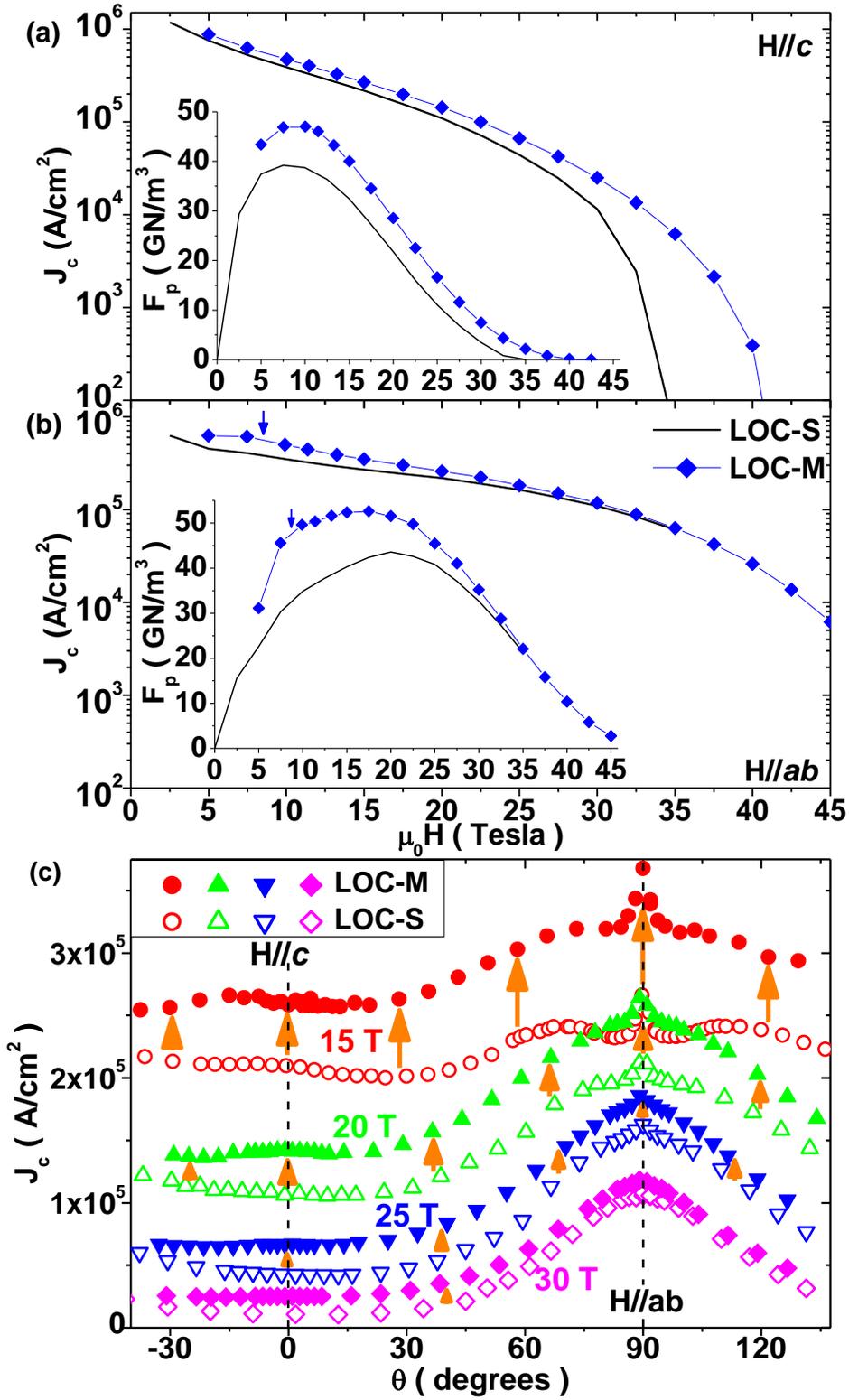

Figure 8